\documentclass[12pt]{article}
\usepackage{epsfig}

\newcommand{\pfrac}[2]{\left(\frac{#1}{#2}\right)}
\newcommand{\GeV}{{\rm\,GeV}}

\newcommand{\al}{\alpha} 
\newcommand{\als}{\alpha_s} 
\newcommand{\bea}{\begin{eqnarray}}
\newcommand{\eea}{\end{eqnarray}}

\newcommand{\msbar}{\overline{\rm MS}}
\newcommand{\ice}[1]{\relax}
\newcommand{\be}{\begin{equation}}
\newcommand{\ee}{\end{equation}}

\begin{document}
\thispagestyle{empty}
\begin{flushright}
MZ-TH/01-13\\
April 2001
\end{flushright}
\vspace{2.5cm}
\begin{center}
{\Large\bf Spectrality, coupling constant analyticity \\[7pt]
and the renormalization group}\\[19pt]
{\large A.A.~Pivovarov}\\[9pt]
Institut f\"ur Physik der Johannes-Gutenberg-Universit\"at,\\[3pt]
Staudinger Weg 7, 55099 Mainz, Germany\\[7pt]
and\\
Institute for Nuclear Research of the\\[3pt]
Russian Academy of Sciences, Moscow 117312
\end{center}
\vspace{1cm}
\begin{abstract}\noindent
Analytic structure in the strong coupling constant
that emerges for some observables in QCD after duality averaging 
of renormalization group improved amplitudes is discussed.
It is shown that perturbation theory calculations are 
justified for the proper observables related to the two-point correlators
of hadronic currents the analytic properties of which 
are well-established. A particular case of gluonic current
correlators is discussed in detail.
\end{abstract}

\vspace{3pt}

\vspace{3pt}

\newpage
General properties of quantum field theory 
impose strong constraints on model building for elementary 
particle phenomenology.
Symmetry properties of interactions 
lie in the foundation of the standard model \cite{SM}, 
causality leads to restrictions on the analytic structure 
of scattering amplitudes as functions of energy,
the freedom of redefinition of ultraviolet subtraction procedures in
renormalizable theories leads to the renormalization group 
invariance which is a basic property of theoretical quantities 
corresponding to physical observables \cite{RG}.
While such general properties are supposed to be valid in a 
full theory there is little
known about the very existence of realistic nontrivial 
quantum field models -- only some simplified examples
(mainly in two-dimensional space-time) have been
considered (e.g.~\cite{twodim}).
The realistic four-dimensional models suitable for particle
phenomenology are analyzed within perturbation theory in the
coupling and only first few terms of perturbative expansions
are usually available. Some general results on asymptotic
behavior at large orders of perturbation theory   
are obtained by the steepest descent method for the functional
integral determining the generating functional for Green's functions.
The results are known in some models of quantum field theory 
where classical solutions of equations of motion 
were found~\cite{lipatov}. The classical solutions of field
equations are also known 
in nonabelian gauge theories~\cite{insta}
that provides the appropriate 
saddle-point configurations for the steepest descent method of evaluating 
the functional integrals~\cite{thooft}
and allows for deeper understanding
the ground state structure in these models~\cite{complvac}.
Besides the steepest descent methods for evaluating 
functional integrals the all-order perturbation theory results 
are also discussed using 
a particular way of resumming some special
subsets of perturbation theory diagrams~\cite{i1nc,qed}.

At present the problem of evaluating the high-order perturbation theory
contributions becomes a practical issue for high-precision tests of 
the standard model and new physics search as the accuracy
of experimental data 
improves \cite{PDG}. It is most important in  perturbative
QCD because the strong coupling constant 
$\al_s$ is numerically large. Since the perturbation theory expansion 
in $\al_s$ is, in general,
asymptotic a resummation of all-order terms gives a possible way to
improve the accuracy of theoretical predictions.
An example of the infinite resummation 
of perturbation theory diagrams is 
an account for the Coulomb interaction for the processes of heavy quark 
production near the threshold \cite{khoze}
that allowed for an essential improvement in the
description of top-antitop production \cite{Hoang:pptt}.
Note that this resummation does not really include the strong
coupling regime of QCD.
For light quarks and massless gluons with a genuine strong
interaction in the infrared domain
there is no successful recipe of resumming the 
subsets of perturbation theory diagrams that could lead
to the description of observables in terms of physical hadrons
\cite{hoosch}. 
To deal with the region of strong coupling in 
the low-energy hadron phenomenology one exploits an idea
of averaging over some energy range. It is assumed that the
theoretical predictions for averaged
quantities obtained with perturbation theory 
in the strong coupling constant in terms 
of quark-gluon degrees of freedom 
can be well confronted with experimental data measured in terms of 
observed hadrons. This assumption is known as duality concept 
(e.g. \cite{locdual2:locdual}).
While the duality assumption is a real base for using perturbation
theory in the low-energy hadron phenomenology it is, 
however, difficult to quantitatively control the accuracy of this assumption
in concrete applications. The most advanced quantitative study of 
the validity of duality concept is for two-point correlators of hadronic
currents because of their simple analytic properties in momentum. 
The quality of the perturbation theory series for two-point correlators
can be essentially improved by the renormalization group resummation
that is an efficient 
tool of calculating various asymptotics of the Green's functions
and related to the freedom of performing ultraviolet subtractions 
that leads to a possibility of redefinition of the coupling.
The technical way to implement the renormalization group
improvement of perturbation theory series
is to use a running coupling normalized in the vicinity
of a physical scale of the process in question.
Such a choice of normalization for the coupling 
allows one to resum big logarithms related to the difference of scales
in all orders of perturbation theory.
Because of the final average over the energy interval as duality 
requires one has a choice whether the renormalization
group improvement
should be done before or after averaging. In general,
these two operations -- duality averaging and renormalization
group improvement -- do not commute.
Performing renormalization group improvement 
before the final averaging allows one to resum a
lot of regular corrections relevant to running only; one can consider
this procedure as a determination 
of a proper scale for the averaged observables.
The technique of renormalization group improvement 
for two-point correlators before the final 
averaging necessary for physical observables
is known as the contour-improved perturbation theory 
and is especially important at low energies where the QCD coupling
constant  is large and higher order perturbation theory 
terms can be numerically important:
they can change the results of finite-order perturbation theory by 
an amount comparable with experimental precision \cite{Pivtau}.
The precision of  present experimental data on $\tau$ lepton decays,
for instance, suffices for distinguishing the results of contour-improved 
and finite order perturbation theory \cite{exp}.

An account for running in perturbation theory
by using the renormalization group improvement under integration sign
is close in spirit to the formulation of calculational scheme for
the Green's functions within Schwinger-Dyson equations (skeleton expansion).
Within Schwinger-Dyson formulation of perturbation theory
one can use for the irreducible vertices
which constitute the building 
blocks of the integral equations 
either finite-order perturbation theory or 
renormalization group improved one.
The Schwinger-Dyson technique was intensively used for determination 
of the fermion propagator beyond the QCD perturbation theory
approximation in relation to 
the problem of mass generation in massless theories 
and spontaneous symmetry breaking \cite{fermprop}. 
It is known that reiteration of running into loops can be infrared 
dangerous (just to have an idea what happens 
one can think of perturbation theory expansions in terms of a bare
coupling in dimensional regularization and compare the results to 
the situation in superrenormalizable theories with a dimensional
coupling constant). The reason is that the renormalization group 
summation is 
applicable to Green's functions at some values of momenta 
while the asymptotic behavior is determined by 
performing an analytic continuation which is sometimes implicit.
Therefore, analytic properties of amplitudes in the whole complex
plane of momenta in finite-order
perturbation theory can differ from those after renormalization group
improvement. This difference of analytic properties 
can lead to some singularities when perturbative 
renormalization group running is extended to area 
where perturbation theory is not valid \cite{mull}.
For asymptotically free QCD in the leading order of running 
the situation was discussed in \cite{zakh}.

In the present paper we discuss the resummation of 
effects of running    
on the example of two-point correlators of hadronic currents.
The two-point correlators are about the simplest Green's functions 
and have well-established analytic properties in momenta.
Two-point correlators are important for phenomenology, they are
relevant for describing the processes of   
$e^+e^-$-annihilation into hadrons and/or $\tau$-lepton hadron decays
\cite{pichrevChetyrkin:1998ej}.
Note that the correlators of gauge invariant currents built 
from gluonic operators describe a spectrum of glueballs,
the experimental observation of which would give a strong 
additional support for QCD as theory of hadrons.
Gluonic current correlators is an actual choice for the analysis 
in the present paper. 

We first discuss some generalities. The correlator 
of a hadronic current $j(x)$ has the form
\begin{equation}\label{correlator}
i\int\langle Tj(x)j^\dagger(0)\rangle
  e^{iqx}dx=\Pi(q^2)
\end{equation}
where $\Pi(q^2)$ is an invariant scalar function.
Analytic properties of the function
$\Pi(q^2)$ in the variable $q^2$ are fixed 
by a dispersion relation 
(K\"allen - Lehmann, or spectral,  representation)
\begin{equation}
\label{disp}
\Pi(q^2)=\int\frac{\rho(s)ds}{s-q^2}+{\rm subtractions} 
\end{equation}
where the spectral density $\rho(s)$ is determined by a sum over
the states of the theory  (e.g. \cite{Itzykson})
and ultraviolet subtractions is a polynomial 
in $q^2$. The spectrum of the correlator in eq.~(\ref{correlator}), or 
the support of the function $\rho(s)$ from eq.~(\ref{disp}),
is determined by singularities of the function $\Pi(q^2)$ in the complex 
$q^2$ plane. The 
spectral density $\rho(s)$ is then given by the discontinuity  
of the function $\Pi(q^2)$ across the spectrum
\begin{equation}
\label{rho}
\rho(s)=\frac1{2\pi i}(\Pi(s+i0)-\Pi(s-i0)),\qquad s\in [{\rm
spectrum}]\, .
\end{equation}
In QCD with massless quarks and gluons
a general assumption about the spectrum (spectrality condition) 
is $s\ge 0$ or $[{\rm spectrum}]=[0,\infty]$.
This assumption is based on the Fock representation for the states 
in terms of massless quarks and gluons (e.g. \cite{Itzykson}). 
Note that this is an assumption and, 
in fact, analytic properties of $\Pi(q^2)$ and, therefore, the
support of the spectral density $\rho(s)$ depend on interaction. 
The dependence of the spectrum on interaction can readily be seen 
in the example of heavy charged particles with 
Coulomb interaction. For a pair of heavy particles with masses 
$m_1$ and $m_2$ one would expect the spectrum start at 
the threshold $s_{\rm thr}=(m_1+m_2)^2$.
However, if the Coulomb interaction is present it is true only for 
the repulsive interaction while the attractive interaction leads to 
the appearance of Coulombic poles below the threshold.
In QCD the shape of the spectrum near the heavy quark
threshold depends also on definition of the masses used to describe heavy
quarks and other details of the interaction (e.g. theoretical 
spectrum can be different at different orders 
of perturbation theory \cite{last}).
Such a situation is well known also from the analysis of simplified models 
\cite{manymodels}.
Thus, the theoretical spectrum of a hadronic correlator 
is a dynamical quantity and constraints on the support of the
spectral density 
coming from kinematical considerations based on values of masses of asymptotic
states are not always valid in the full theory.

In asymptotically free QCD the function
$\Pi(q^2)$ is computable theoretically in
Euclidean domain (sufficiently far from the positive semiaxis
$q^2>0$) that allows one to find theoretical predictions for
observables.  Still,
to extract a theoretical prediction for the spectral density 
$\rho(s)$ from the function $\Pi(q^2)$ is
not straightforward. The point is that $\Pi(q^2)$ is only known as 
a perturbation theory expansion at large Euclidean 
$q^2$ while $\rho(s)$ is
given by a discontinuity across singularities of $\Pi(q^2)$
in the complex $q^2$ plane. However, the perturbation theory
calculation of the function 
$\Pi(q^2)$ is not justified near its singularities. Therefore, 
the analytic continuation in the complex $q^2$ plane to the vicinity
of positive semiaxis and into infrared region is necessary. 
The analytic
continuation is an incorrectly set operation, i.e. small errors of the
initial function $\Pi(q^2)$ at Euclidean points 
can produce large errors in $\rho(s)$. This
instability is especially important for a theoretical evaluation of
$\rho(s)$ at low energy. 
The problem of performing an analytic continuation
can be also reformulated in
the language of integral equations since the dispersion relation in
eq.~(\ref{disp}) gives the
correlation function $\Pi(q^2)$ as an integral transformation
of the spectral density $\rho(s)$. The
relation~(\ref{disp}) is a Fredholm integral
equation of the second kind which is known to lead to an incorrectly
set problem.  Thus, the errors of $\rho(s)$ (as a solution of 
equation~(\ref{disp})) are not
continuously related to the errors of $\Pi(q^2)$ 
(as initial data of equation~(\ref{disp})) 
and can be large even if errors of $\Pi(q^2)$ 
in Euclidean points are sufficiently small.  
The general procedure of constructing the approximate solutions to 
incorrectly set problems was suggested by Tikhonov and
is known as regularization. Averaging the spectral density over a
finite energy interval (sum rules) can be considered as a particular
realization of Tikhonov's regularization procedure.
One wants to theoretically study the function $\rho(s)$ at low energy 
because its experimental
counterpart -- the hadronic spectral density 
$\rho^{\rm had}(s)$ -- can be directly measured at low energy
with high precision. Thus, while a pointwise
description of the spectral density $\rho(s)$ at low energy
is beyond perturbation theory, the appropriate quantities to
analyze theoretically in perturbative 
QCD are the moments or integrals of $\rho(s)$
with a set of weight functions. This is a manifestation of 
the fact that the theoretical spectral density 
is, in general, a distribution rather 
than a continuous function of energy. 

The moments of the spectral density  $\rho(s)$
over a finite energy interval are defined by the relation
\begin{equation}
\label{intmom}	
M_n=(n+1)\int^{s_0}\rho(s)\frac{s^nds}{(s_0)^{n+1}}\, .
\end{equation}
The factor $(n+1)$ in the definition of the moments is chosen to have
all contributions of the leading order of perturbation theory 
uniformly normalized to unity.
Equivalently one can say that all measures defined on
the interval $[0,s_0]$
\begin{equation}
\label{measure}
(n+1)\frac{s^n}{s_0^{n+1}}ds=d\pfrac{s}{s_0}^{n+1}
\end{equation}
are normalized to unity.
Note that the accuracy of perturbation theory evaluation of a
given moment depends on a particular weight function.

With the dispersion relation given in eq.~(\ref{disp})
one can rewrite moments in eq.~(\ref{intmom})	
as integrals over a contour in the complex $q^2$ plane 
\cite{cont}. For 
practical calculations of moments a convenient contour is a
circle with the radius $s_0$ though the results are independent of
the shape of the contour when it is deformed in the analyticity
domain of the correlator. The contour representation of the 
moments reads
\begin{eqnarray}
\label{momcirc}
M_{n}&=&(n+1)\frac{(-1)^n}{2\pi i}\oint_{|z|=s_0}
  \Pi(z) (z/s_0)^n dz/s_0\ \nonumber\\
&=&(n+1)\frac{(-1)^n}{2\pi i}\oint_{|z|=1}
  \Pi(s_0z)z^n dz\ \nonumber\\
  &=&(n+1)\frac{(-1)^n}{2\pi}\int_{-\pi}^\pi
  \Pi(s_0e^{i\varphi}) e^{i(n+1)\varphi}d\varphi\, .
\end{eqnarray}
Note that the moments on the circle
as given in eq.~(\ref{momcirc}) are just Fourier coefficients of correlation
functions that allows one to use a well-developed mathematical
technique of Fourier analysis to study them.

Theoretical calculations of the moments are usually performed within 
operator product expansion (OPE) 
for the correlation function $\Pi(q^2)$~\cite{wilsonOPE,politzer,svz}. 
The OPE expression for the correlator contains a perturbation theory 
part and power corrections. 
The perturbation theory part can be further improved
using renormalization group summation.
In this paper we consider only the perturbation theory part of
the theoretical correlator, or $\Pi(q^2)$-function, 
for analyzing the moments. If the renormalization group
improved $\Pi(q^2)$ is used under integration sign for the moments 
this means a resummation of the effects of
running~\cite{Pivtau}. This technique was used for 
tau decay analysis~\cite{Pivtau,DP}.
Power corrections within OPE -- nonperturbative terms --
appear by prescribing 
the nonvanishing vacuum expectation values to 
the local operators of higher dimensionality~\cite{svz}.
The contributions of these terms into the moments can be
found with Cauchy theorem (e.g.~\cite{FESR1}).
At present the qualitative change in the phenomenology of sum rules 
is that high-order perturbation theory terms for hadronic
correlators are known in various hadronic channels
and the experimental value for the 
strong coupling constant is larger than that of the original papers 
therefore perturbation theory corrections are important numerically. 
It was already noticed that in some channels the 
perturbation theory corrections
can numerically dominate over the power corrections 
that makes the study of perturbation theory corrections 
important for the present phenomenology~\cite{spinn}.

As a concrete example we take a correlator of the gluonic current
$G^2=G_{\mu\nu}^a G_{\mu\nu}^a$ where $G_{\mu\nu}^a$
is a strength tensor
of the gluon field. To the leading order of perturbation theory 
the renormalization group invariant expression for the current can be
chosen in the form
\be
j_G=\al_s G^2
\ee
where $\al_s$ is the strong coupling constant of QCD. 
This current is related to the trace of the QCD energy-momentum
tensor $\theta_\mu^\mu$ 
and can serve as interpolating operators for glueballs.
The full renormalization group invariant expression for
$\theta_\mu^\mu$ 
in QCD with massless quarks is 
$(\beta(\al_s)/2\al_s) G^2$
where $\beta(\al_s)$ is the QCD $\beta$-function; 
this is not important for us in the following.
The correlator reads
\be
\label{corrG}
\frac{\pi^2}{2} 
i\int\langle Tj_G(x)j_G^\dagger(0)\rangle  e^{iqx}dx
=q^4 \Pi_G(q^2) \, .
\ee
Note that a kinematical factor $q^4$ is removed from the definition of
the function $\Pi_G(q^2)$ which is justified within perturbation theory.
The Adler's function
\begin{equation}
\label{adler}
D_G(Q^2)=-Q^2\frac{d}{dQ^2}\Pi_G(Q^2)\, , \quad Q^2=-q^2
\end{equation}
has a simple form at the leading order of perturbation theory
\begin{equation}
\label{main}
D_G(Q^2)=\als(Q^2)^2(1+O(\als))\, .
\end{equation}
A theoretical prediction for the function $D_G(Q^2)$ 
has been calculated up to the third order of perturbation 
theory~\cite{kksch,OkadaChetyrkin:Higgs}. 
Our main aim is to take into account the effects of running of the coupling
for evaluating the moments of the spectral density, 
therefore, the introduction of an effective charge is convenient 
\cite{effsch}. Indeed, 
high-order corrections can be accounted for by introducing 
the effective charge $\alpha_G(Q^2)$
in all orders of perturbation theory by the relation \cite{kksch}
\begin{equation}
\label{effG}
D_G(Q^2)=-Q^2\frac{d}{dQ^2}\Pi_G(Q^2)=\alpha_G(Q^2)^2\, .
\end{equation}
The effective strong coupling $\alpha_G(Q^2)$
obeys the renormalization group equation
\begin{equation}
\label{RGfull}
Q^2\frac{d}{dQ^2}\frac{\alpha_G(Q^2)}\pi=\beta (\frac{\alpha_G(Q^2)}{\pi}) 
\end{equation}
with 
\begin{equation}
\label{betaF}
\beta (a)=-a^2
\left(\beta_0+\beta_1 a+\beta_2^G a^2+\beta_3^G a^3\right)+O(a^6)\, .
\end{equation}
 First two coefficients of the $\beta$-function are scheme
independent,
the higher order terms $\beta_2^G,\beta_3^G$ depend on the effective
charge definition in eq.~(\ref{effG}).
In QCD with $n_f$ light quark flavors one has 
\be
\beta_0=\frac{1}{4}\left(11-\frac{2}{3}n_f\right)\, .
\ee
 For our purpose it suffices to use only the leading order
running that contains all essential features of the whole phenomenon. 
Effects due to
higher order corrections of the $\beta$-function are small and do not
change the basic picture, 
slightly affecting the values of the moments numerically~\cite{groote}.
Thus, we consider the leading order renormalization group
equation for the effective charge
\begin{equation}
\label{RG0}
Q^2\frac{d}{dQ^2}\frac{\alpha_G(Q^2)}\pi=-\beta_0\pfrac{\alpha_G(Q^2)}\pi^2. 
\end{equation}
The renormalization group resummed correlation function reads
\begin{equation}
\label{piRG}
\Pi_G(Q^2)=\frac{\pi}{\beta_0}\alpha_G(Q^2)
+{\rm subtractions}
\end{equation}
where
\begin{equation}
\label{alpfunc}
\alpha_G(Q^2)=\frac{\alpha_0}{1+(\beta_0\alpha_0/\pi)
  \ln(Q^2/s_0)}
\end{equation}
with $\alpha_0=\alpha_G(s_0)$. 
Note that for the process of $e^+e^-$ annihilation into hadrons
the corresponding 
renormalization group resummed correlation function reads 
\begin{equation}
\label{ee}
\Pi_{e^+e^-}(Q^2)
=\ln\pfrac{\mu^2}{Q^2}
+\frac1{\beta_0}\ln\pfrac{\alpha_{e^+e^-}(Q^2)}\pi
+{\rm subtractions}
\end{equation}
with the first term being a
parton contribution independent of $\alpha_s$.
Setting  $Q^2=s_0e^{i\varphi}$ on the
contour one obtains an explicit expression for the correlator 
as a function of the angle ${\varphi}$
\begin{equation}
\label{piGphi}
\Pi_G(s_0e^{i\varphi})=\frac{\pi}{\beta_0}
\frac{\al_0}{1+i{\beta_0}\alpha_0\varphi/\pi}
+\mbox{subtractions}
\end{equation}
With an explicit expression for the function
$\Pi_G(z)$ from eqs. (\ref{piRG},\ref{piGphi})
the analysis of the moments $M_{n}$ is straightforward.
The explicit expression for the moments written through the contour
representation reads
\be
\label{momG}
M_{n}=(n+1)\frac{(-1)^n}{2\pi}\int_{-\pi}^\pi
\frac{\pi}{\beta_0}\frac{\al_0}{1+i{\beta_0}\alpha_0\varphi/\pi}
e^{i(n+1)\varphi}d\varphi\, .
\ee
Eq.~(\ref{momG}) is a basic relation
for further study. Note that the form of the representation
in eq.~(\ref{momG}) is rather general and 
gives a basis for other applications:
higher powers of the running coupling $\alpha_s$ can be easily generated.

Let us discuss the above expressions in some detail.
The main feature of the contour representation for the moments 
is that everything is explicit as it is in finite-order perturbation
theory. After formulating the particular way of
resummation for the moments, i.e. by defining them on the contour,
there is no ambiguity in these quantities 
(they are not given by series in $\al_s$ but
by close formulae). 
Therefore, the moments are explicit functions of $\al_0$
that can be rigorously studied.
One should, however, remember that a particular
definition of the moments on the contour has been used.

Expanding eq.~(\ref{momG})
in $\alpha_0$ one reproduces all results of finite-order perturbation
theory (e.g.~\cite{Korner:2000ef}).
Indeed, expanding the function $\Pi_G(Q^2)$ from 
eq.~(\ref{piRG}) one finds 
\be
\label{piFOPT}
\Pi_G(Q^2)=\frac{\pi}{\beta_0}\alpha_0
\left\{ 1
+\beta_0\frac{\alpha_0}{\pi}\ln(\frac{s_0}{Q^2})
+\beta_0^2\left(\frac{\alpha_0}{\pi}\right)^2\ln^2(\frac{s_0}{Q^2})
+O(\alpha_0^3)\right\}
+{\rm subtractions}\, .
\ee
The first term ($Q^2$ independent) can be added to subtractions. 
Then, finally one has 
\be
\label{piFOperturbation theoryfin}
\Pi_G(Q^2)=\alpha_0^2\ln\left(\frac{s_0}{Q^2}\right)
+\beta_0\frac{\alpha_0^3}{\pi}\ln^2\left(\frac{s_0}{Q^2}\right)
+O(\alpha_0^4)+{\rm subtractions}\, .
\ee
While the expansion of the integrand in
eq.~(\ref{momG}) in $\alpha_0$ 
with further integration gives nothing new in
comparison with the finite-order perturbation theory analysis, 
new features appear if one retains a resummed expression 
under integration sign.

The moments in eq.~(\ref{momG}) are expandable in a convergent
series in $\alpha_0$ for $\beta_0\alpha_0<1$. 
The existence of a finite radius of convergence in the complex
$\alpha_0$ plane within the contour 
technique of resummation for the moments 
is a general feature that persists 
for the running with the high-order perturbative 
$\beta$-function. However, in QCD the convergence radius in 
$\alpha_0$ decreases when higher orders of the $\beta$-function
are included~\cite{groote}. 
Thus, the explicit result eq.~(\ref{momG})
allows for an analytic continuation in the complex $\alpha_0$ plane
leading to the functions $M_n(\alpha_0)$ which are analytic
in $\alpha_0$ at the origin, i.e. near the point $\alpha_0=0$. 
This sounds a bit unusual as one implicitly assumes
that perturbation theory objects should have an
essential singularity in $\alpha_0$ at the origin usually a cut along
the negative semiaxis (e.g.~\cite{khuri}). Note that the moments of the
heavy quark production with infinite resummation of Coulomb interaction are
also given by convergent series in $\al_s$ (the explicit result 
at the leading order of perturbation theory is presented 
in~\cite{ppCoulmom}).
The exact expression
given in eq.~(\ref{momG}) without expansion in $\alpha_0$ provides one
with an analytic continuation beyond the convergence radius
even when $\al_0$ lies outside the convergence circle. 

Looking at 
eqs.~(\ref{disp},\ref{rho},\ref{piRG},\ref{alpfunc},\ref{piGphi}) 
one notices that analytic properties in the variable $q^2$
declared for a general function $\Pi(q^2)$ built from massless fields
in eqs.~(\ref{disp},\ref{rho}) differ from that  
of the explicit result given in 
eqs.~(\ref{piRG},\ref{alpfunc},\ref{piGphi}): the explicit 
renormalization group improved expression $\Pi_G(q^2)$
has a pole in the Euclidean region of $q^2$ which is supposed to be
the analyticity region from general assumptions about
the spectrum. This is an important feature to notice:
a concrete approximation $\Pi_G(q^2)$ in eq.~(\ref{piRG}) has
different analytic properties in the whole complex $q^2$ plane than
it is declared by general requirements. Contrary to the resummed
expression given in eq.~(\ref{piRG}), 
at any finite order of perturbation theory given in eq.~(\ref{piFOPT})
one has only powers of logarithms
that have correct analytic properties in the variable $q^2$
-- a cut along the positive semiaxis.
It is just a consistency feature -- 
finite-order perturbation theory is a (trivial) example of the
model of quantum field theory where all general requirements are valid.
Thus, the renormalization group resummation 
for the hadronic correlator in asymptotically free QCD
can change its analytic structure in the infrared region as compared to 
the finite-order perturbation theory approximation.
In the leading order of the running in QCD 
a (Landau) pole is usually generated.
This pole is included into the definition of the moments 
in eqs.~(\ref{momcirc},\ref{momG})
because one encircles the origin with a
large contour. There is no other possibility to work consistently
in perturbation theory because the infrared 
region is completely nonperturbative
and one is not allowed to move the integration 
contour to that region.
The requirement of integrating only along the positive axis
is an external constraint on the theory 
rather than its attribute.
It cannot be realized in perturbation theory -- the integration
contour should go sufficiently far
from the infrared region which is a requirement of 
the applicability of perturbation theory approximations. 
Note that if $s_0$ is not large enough in order the circular contour
includes all infrared singularities the contour should be 
deformed to do so. To give the results for the moments that are 
justified in perturbation theory (at least formally)
the integration contour should be chosen such that no
singularity incompatible with general requirements 
lies outside it in the complex $q^2$ plane.  

After defining the moments properly
(written as eq.~(\ref{momG}), for instance) 
the practical calculation of explicit functions $M_n(\al_0)$
can be done in different ways. 
Technically, one can shrink the integration contour back 
to singularities of $\Pi_G(q^2)$ which is a uniquely defined
mathematical operation for the explicit function 
$\Pi_G(q^2)$ in the complex $q^2$ plane.
Then one discovers a pole which is
a pure computational fact without any meaning for the structure of
the perturbation theory at high orders. 
The perturbation theory moments are constructed at high energies
and cannot decipher the point-by-point structure of the spectrum 
in the infrared region
(or singularities of $\Pi_G(q^2)$ at small $q^2$) --
they just give a contribution from this region as it is seen 
from large energies (on the contour).
If a high-order $\beta$-function is used for the renormalization
group improvement of the correlator 
then the structure of singularities in the
infrared region can drastically change \cite{renRS}.
However, this has little
effect on the moments -- they develop
some small perturbation theory corrections 
independent of a particular structure
of the correlation function in the infrared region obtained as 
a perturbation theory approximation. 
Of course, the parameter $s_0$ should be
sufficiently large in order the perturbation theory expansion in 
the coupling $\al_0$ would be justified. A discussion of 
the pointwise behavior of $\Pi(q^2)$ in the infrared region 
is beyond the scope of perturbation theory. 
Note that the possibility to restore moments as exact functions 
of the coupling from their (asymptotic) perturbation
theory series depends on the behavior of $\Pi(q^2)$ in the infrared
region.

Still some convenient 
representations of eq.~(\ref{momG}) are worth studying
for practical calculations. 
Let us first consider the leading order moment $M_0(\al_0)$
that reads
\be
\label{momG00}
M_{0}=\frac{1}{2\pi}\int_{-\pi}^\pi
\frac{\pi}{\beta_0}\frac{\al_0}{1+i{\beta_0}\alpha_0\varphi/\pi}
e^{i\varphi}d\varphi
=\frac{\al_0}{2\beta_0}\int_{-\pi}^\pi
\frac{e^{i\varphi}d\varphi}{1+i{\beta_0}\alpha_0\varphi/\pi}\, .
\ee
With the expression~(\ref{momG00})
given one can work out an efficient computation technique 
for it.  
In applications the moments are usually computed numerically 
\cite{Pivtau,DP,groote}. For a general analysis
one can consider also analytical computations of the moments
in various limits.
Integrals of the type~(\ref{momG00}) are related to the exponential
integral function and have been well studied~\cite{bookY}. 
Formally one can use a convergent series in $\al_0$
but if an experimental value of $\al_0$ is larger than the 
convergence radius then the expansion in  $\al_0$ 
is of no use and an analytic continuation of the function given
by the series in $\alpha_0$ beyond
the convergence radius is necessary.
Let us look at this issue closer.
The convergence radius of the function $M_0(\al_0)$ in the complex
$\al_0$ plane
for the leading order $\beta$-function is given by 
$|\alpha_0|<1/\beta_0$. For a full perturbative $\beta$-function up to
the fourth order in the $\msbar$ scheme it is smaller \cite{groote}.
In a realistic case of $\tau$ decays, for instance, 
$s_0=M_\tau^2=(1.777~\GeV)^2$ 
and $\beta_0=9/4$ that leads to 
\be
\alpha_0\equiv \alpha_0(s_0=M_\tau^2=(1.777~\GeV)^2)
<\frac{4}{9}=0.44\ldots
\ee
i.e. the experimental value of the coupling $\alpha_0\approx 0.3$
\cite{kr3} 
lies rather close to the boundary of the convergence circle. 
Taking the scale $s_0$ for the moments 
smaller than the squared $\tau$ lepton mass $M_\tau^2$
one can get the value of $\alpha_0$ lying outside the convergence circle.
The convergent power series may not be the best way of
computing the moments for such numerical values of the coupling constant. 
The more efficient approximation can be obtained by constructing an
asymptotic expansion for the zero moment.
Integrating by parts one finds 
\be
\label{momcircBas2}
M_{0}=\frac{\alpha_0^2}{1+\beta_0^2\alpha_0^2}+
\frac{\alpha_0^2}{2\pi}\int_{-\pi}^\pi
  \frac{e^{i\varphi}d\varphi}{(1+i\beta_0\alpha_0\varphi/\pi)^{2}}\, .
\ee
Here the first term gives a perturbation theory expression for 
the spectral density at $s_0$ with all corrections due to
analytic continuation resummed (so called $\pi^2$ corrections)
\cite{picorr}.
This contribution can be obtained from the leading order running
by retaining the highest power of $\pi$ at every order 
of perturbation theory.
It also corresponds to the calculation of the moments on the cut
through the boundary value of the perturbation theory spectrum 
\cite{Korner:2000ef}. Further integration by parts gives
\begin{eqnarray}
\label{mom0byparts}
M_{0}&=&\frac{\alpha_0^2}{1+\beta_0^2\alpha_0^2}
  + \frac{\alpha_0^2}{\pi}
\sum_{j=2}^{n}(j-1)!\pfrac{\beta_0\alpha_0}{\pi}^{j-2}
  \frac{\sin\{j\arctan(\beta_0\alpha_0)\}}
{(1+\beta_0^2\alpha_0^2)^{j/2}}
\nonumber\\&&
+n! \frac{\alpha_0^2}{2\pi}\pfrac{\beta_0\alpha_0}\pi^{n-1}\int_{-\pi}^\pi
\frac{e^{i\varphi}d\varphi}{(1+i\beta_0\alpha_0\varphi/\pi)^{n+1}}\, .
\end{eqnarray}
This result can be obtained using the recurrence relation
\be
\label{rec}
\frac{1}{2}\int_{-\pi}^\pi
  \frac{e^{i\varphi}d\varphi}{(1+i\beta_0\alpha_0\varphi/\pi)^{k}}
=\frac{\sin(k\chi)}{r^k}
+k\pfrac{\beta_0\alpha_0}{\pi}\frac{1}{2}\int_{-\pi}^\pi
  \frac{e^{i\varphi}d\varphi}{(1+i\beta_0\alpha_0\varphi/\pi)^{k+1}}
\ee
with quantities $r$ and $\chi$ defined by
\begin{equation}
1+i\beta_0\alpha_0=re^{i\chi},\qquad
r=\sqrt{1+\beta_0^2\alpha_0^2},\qquad
\chi=\arctan(\beta_0\alpha_0)\, .
\end{equation}

Retaining only leading powers of $\alpha_0$ at every
order of the expansion~(\ref{mom0byparts})
one recovers an asymptotic series often discussed in the literature. 
Indeed, taking only the leading asymptotics of every term 
in eq.~(\ref{mom0byparts}) one finds 
\be
\label{wrong}
M_{0}^{\rm leading~asym}=\alpha_0^2\left(
1+2\beta_0\frac{\alpha_0}{\pi}+\ldots+
(n+1)!\beta_0^n \pfrac{\alpha_0}{\pi}^n\right)+O(\alpha_0^{n+3})\, .
\ee
The expansion~(\ref{wrong}) shows a nonalternating 
factorial growth of the coefficients 
that leads to a Borel nonsummable asymptotic series
\cite{factor1factor2}. The approximation~(\ref{wrong}) for 
the expansion~(\ref{mom0byparts}) is not accurate. 
Note that Borel summation (with some recipe for treating nonsummable
singularities) of the leading asymptotics~(\ref{wrong})
cannot restore the exact 
function~(\ref{momG00}) from some general principles.  

The representation~(\ref{mom0byparts})
gives an efficient way to numerically compute the result (\ref{momG00}),
it represents an asymptotic expansion of the function $M_0(\al_0)$
which is analytic at the origin $\al_0=0$. 
It is known that an asymptotic expansion of a function can be
more efficient for its numerical evaluation than a
convergent series even inside the convergence circle, 
it also gives an efficient way for the calculation outside 
the convergence circle (not too far though).
One can see that the result~(\ref{mom0byparts}) is an efficient asymptotic
expansion which can give a better accuracy than a direct power series 
expansion in $\alpha_0$ for some $\alpha_0$ and $n$. 
Note that when the analytic structure of the function is known,
or a concise expression for the function is given as in eq.~(\ref{momG00}),
the asymptotic expansions which converge fast for the first few terms 
are more useful for practical calculations than formal convergent
series that require
many terms for getting a reasonable numerical accuracy~\cite{hardy}. 
Still one is left with the residual term which is represented by the
integral in eq.~(\ref{mom0byparts}).
In practical calculations one can simply neglect it.
However, in some cases one can do better than that.
By extending the integration range in the variable 
$\varphi$ from $-\infty$ to $+\infty$ the integral over $\varphi$ 
can be readily computed
\be
\label{infint}
n!\pfrac{\beta_0\alpha_0}{\pi}^{n-1}\int_{-\infty}^{\infty}
\frac{e^{i\varphi}d\varphi}{(1+i\beta_0\alpha_0\varphi/\pi)^{n+1}}
=2\pi \pfrac{\pi}{\beta_0\alpha_0}^2
e^{-\frac{\pi}{\beta_0\alpha_0}}\, .
\ee
Using the decomposition
\be
\int_{-\pi}^\pi d\varphi=\int_{-\infty}^{\infty}d\varphi
-\left(\int_{-\infty}^{-\pi}d\varphi+\int_{\pi}^{\infty}d\varphi\right)
\ee
one can write
\begin{eqnarray}
\label{end}
\lefteqn{n!\pfrac{\beta_0\alpha_0}\pi^{n-1}\int_{-\pi}^\pi
\frac{e^{i\varphi}d\varphi}{(1+i\beta_0\alpha_0\varphi/\pi)^{n+1}}
\ }\nonumber\\
&=&2\pi\pfrac{\pi}{\beta_0\alpha_0}^2e^{-\frac{\pi}{\beta_0\alpha_0}}
-n!\pfrac{\beta_0\alpha_0}\pi^{n-1}
\left(\int_{-\infty}^{-\pi}+\int_{\pi}^{\infty}\right)
\frac{e^{i\varphi}d\varphi}{(1+i\beta_0\alpha_0\varphi/\pi)^{n+1}}
\end{eqnarray}
for any $n$. Therefore, the residual term 
in eq.~(\ref{mom0byparts}) is transformed into a sum 
of an explicit nonperturbative term proportional to 
$e^{-\pi/\beta_0 \al_0}$ and the term which can be smaller than the
original residual term for some values of $\alpha_0$ and $n$. 
One has
\begin{eqnarray}
\label{mom0bypartsNONPT}
M_{0}&=&\pfrac{\pi}{\beta_0}^2
e^{-\frac{\pi}{\beta_0\alpha_0}}
+\frac{\alpha_0^2}{1+\beta_0^2\alpha_0^2}
  + \frac{\alpha_0^2}{\pi}
\sum_{j=2}^{n}(j-1)!\pfrac{\beta_0\alpha_0}{\pi}^{j-2}
  \frac{\sin\{j\arctan(\beta_0\alpha_0)\}}
{(1+\beta_0^2\alpha_0^2)^{j/2}}
  \nonumber\\&&
  -n!\frac{\al_0^2}{2\pi}\pfrac{\beta_0\alpha_0}\pi^{n-1}
  \left(\int_{-\infty}^{-\pi}+\int_{\pi}^{\infty}\right)
  \frac{e^{i\varphi}d\varphi}{(1+i\beta_0\alpha_0\varphi/\pi)^{n+1}}\, .
\end{eqnarray}
The explicit nonperturbative term $e^{-\pi/\beta_0 \al_0}$
has appeared in the asymptotic expansion of the
moment~(\ref{momG00}) written in the form of
eq.~(\ref{mom0bypartsNONPT}). 
The difference between the expansions 
in eq.~(\ref{mom0byparts}) and eq.~(\ref{mom0bypartsNONPT})
is not very noticeable, in fact, they are
almost identical up to the residual terms. 
What happened is the change of the residual term. 
Therefore, the choice of the representation for the
moment~(\ref{momG00})
(eq.~(\ref{mom0byparts}) or eq.~(\ref{mom0bypartsNONPT})),
i.e. with or without the 
explicit nonperturbative term $e^{-\pi/\beta_0\al_0}$, is
a question of the choice of a particular form of 
the residual term. 
It can happen that after dropping the residual term 
(which is a common practice in asymptotic series calculations)
the representation in the form of
eq.~(\ref{mom0bypartsNONPT})
is more accurate numerically than that in the form of
eq.~(\ref{mom0byparts}) for some particular values of 
$\al_0$ and $n$. However, 
a quantitative conclusion about the accuracy of 
the asymptotic series representation for a function can only be drawn if 
one has a concise expression for the function as
eq.~(\ref{momG00}) in our case when the explicit form of the
residual term is also known (see also ref.~\cite{modQM}
where a simplified model in quantum mechanics was considered). 
Any conclusions based on the terms of the
series itself (for instance, based on the representation~(\ref{wrong}))
can be not accurate numerically; they  
can also be unjustified in a general sense of analytic behavior 
as one can see from eq.~(\ref{end}).

The above results are valid for any moment $M_{l}$. Namely,  
the recurrence relation can be generalized to read
\be
\label{recmoml}
(l+1)\frac{1}{2}\int_{-\pi}^\pi
  \frac{e^{i(l+1)\varphi}d\varphi}{(1+i\beta_0\alpha_0\varphi/\pi)^{k}}
=\frac{\sin\{k(l+1)\chi\}}{r^k}
+k\pfrac{\beta_0\alpha_0}{\pi}\frac{1}{2}\int_{-\pi}^\pi
\frac{e^{i(l+1)\varphi}d\varphi}{(1+i\beta_0\alpha_0\varphi/\pi)^{k+1}}\, .
\ee
The representation with integration by parts
analogous to one given in eq.~(\ref{mom0byparts})
shows an improvement in the convergence for large $l$ moments
equivalent to the replacement
$\alpha_0\to\alpha_0/l$. This agrees with conclusions drawn from
the analysis of finite-order perturbation theory~\cite{Korner:2000ef}. 
In general, one can also modify the residual term for any moment
$M_n$. In the literature there are also some moments defined 
on the final energy interval with different weight
functions~\cite{malt}; our conclusion can be generalized to that  
moments as well. 

Now we discuss the spectrum of the explicit resummed function $\Pi_G(q^2)$. 
The structure of the spectrum 
in the infrared domain is most interesting. Note that this
part of the spectrum is obtained by the analytic continuation from
the Euclidean region where $\Pi_G(q^2)$ is calculated as a 
perturbation theory expansion to a region where
perturbation theory is not valid that means that 
the structure of the spectrum has no general physical meaning at 
small $s$ pointwise. The spectrum of the explicit
function $\Pi_G(q^2)$ given in eq.~(\ref{piRG}) is a well-defined 
mathematical quantity. It is straightforward to calculate it. 
Using the expression for the
leading order coupling constant in the form 
\begin{equation}
\label{alLambda}
\alpha_G(Q^2)=\frac{\alpha_0}{1+(\beta_0\alpha_0/\pi)
  \ln(Q^2/s_0)}=\frac{\pi}{\beta_0\ln(Q^2/\Lambda_G^2)}
\end{equation}
where
\begin{equation}
\Lambda_G^2=s_0\exp{\left(-\frac{\pi}{\beta_0\alpha_0}\right)}
\end{equation}
one finds
\begin{equation}
\label{piRGlambda}
\Pi_G(Q^2)=\frac{\pi}{\beta_0}\alpha_G(Q^2)+{\rm subtractions}
=\frac{\pi^2}{\beta_0^2\ln(Q^2/\Lambda_G^2)}+{\rm subtractions}\, .
\end{equation}
Therefore, the spectrum (a discontinuity across singularities) reads
\begin{equation}
\label{spectG}
\rho_G(s)=\frac{1}{2\pi i}(\Pi_G(s+i0)-\Pi_G(s-i0))
=\frac{\pi^2}{\beta_0^2}
\left(\Lambda_G^2
\delta(\Lambda_G^2+s)+\theta(s)\frac{1}{\pi^2+\ln^2(s/\Lambda_G^2)}
\right)
\end{equation}
where $\delta(s)$ is a Dirac $\delta$-distribution and 
$\theta(s)$ is a step-distribution.
Explicit functions given in eqs.~(\ref{piRGlambda},\ref{spectG})
satisfy integral equation~(\ref{disp}).
Note that the explicit spectrum in eq.~(\ref{spectG})
contains a contribution 
$\delta(\Lambda_G^2+s)$ corresponding to a pole 
$1/(q^2+\Lambda_G^2)$ of the function $\Pi_G(q^2)$
in the region $q^2<0$
which is supposed to be the analyticity domain of the two-point
correlators from general requirements.
The position of the pole $\Lambda_G^2$ is specific for a given
channel if an effective charge is used. 
The expression for the theoretical spectrum given in
eq.~(\ref{spectG}) can be used in a
mathematical sense for calculating integrals (moments) in 
eq.~(\ref{momG}) (an analogous approach may be used for
the general case in eq.~(\ref{intmom})) but a physical interpretation 
of the spectrum at small $s$ is rather
meaningless because perturbation theory is not applicable at small momenta.

The part of the spectrum on the positive
real axis is a discontinuity of the function $\Pi_G(q^2)$
across the cut~\cite{picorr}. 
It can be conveniently written in the form
\begin{equation}
\label{speccont}
\rho_G^{cont}(s)=\frac{\pi^2}{\beta_0^2}\frac{1}{\pi^2+\ln^2(s/\Lambda^2)}
  =\frac{\alpha(s)^2}{1+\beta_0^2\alpha(s)^2}
\end{equation}
with a function
\begin{equation}
\label{alcut}
\alpha(s)=\frac{\pi}{\beta_0\ln(s/\Lambda_G^2)}\, .
\end{equation}
Note that the function $\alpha(s)$ has a pole on the physical cut
at $s=\Lambda_G^2$.
This is this pole that leads to problems of Borel nonsummability  
in the resummation of the effects of running directly on the cut when
one integrates through the infrared region (cf. eq.~(\ref{wrong})). 
However, the pole of the auxiliary function $\alpha(s)$ from
eq.~(\ref{alcut}) has no physical meaning within perturbation theory.
For instance, the spectral density~(\ref{speccont}) is a smooth function
at this point.
While the spectral density explicitly given in eq.~(\ref{spectG}) 
allows one to compute the moments by the direct integration in 
a pure mathematical 
sense it is not productive to ask whether this spectrum is physical or not
because there is no possibility to answer this question within
perturbation theory. Interpretations of this spectrum at low energies
like specific recipes of resummation for Borel non-summable
series (as in eq.~(\ref{wrong}), for instance, are additional 
assumptions beyond perturbation theory.
 
The continuous part of the spectral density 
in eq.~(\ref{speccont})
can be obtained uniquely from the finite-order perturbation 
theory expansion.
However, the pole remains hidden and cannot be restored from the summation
on the cut if only the discontinuity across the cut along the positive
semiaxis is considered. Note that this is also a situation in heavy
quark physics -- no Coulombic poles can be restored from the summation
on the cut (see discussion in \cite{coulK} in relation to
the precision determination of heavy quark masses).

It is worth stressing again that the moments in eq.~(\ref{momG})
are analytic functions of the coupling $\alpha_0$ at the origin. 
It means that the nonanalytic piece in 
eq.~(\ref{end}) cancels the corresponding part in the residual term. 
Depending on a particular form of the residual term the formal 
analytic structure of the expansion for the moments
in the coupling constant $\al_0$ drastically changes.
This demonstrates a danger of making conclusions about 
infrared power
corrections emerging from the extrapolation of the running to 
the infrared region. Because the infrared region 
is not the perturbation theory domain the formal perturbative 
expansions originating from the integration
over the infrared region can strongly be modified by making 
small changes in perturbation theory quantities like
effective $\beta$-functions \cite{renRS}.  
In practice, or from a phenomenological point of view, the use of
power corrections stemming from 
the infrared modification of perturbation theory 
is difficult to appreciate if high order terms in the
perturbative $\alpha_0$ expansion are taken into account. 
For such observables as the moments of the spectral density one 
cannot distinguish numerically 
high-order perturbation theory corrections from power corrections
(nonperturbative part of the expansion):
the power corrections are numerically hidden by
the high-order perturbation theory corrections.

Thus, for the observables related to 
two-point correlators the problem of resumming the running
effects in perturbation theory is
solved by the contour integration.
We stress that the pole (or any singularity that may occur upon
the formal analytic continuation of the perturbation theory 
expressions into the infrared region)
is inside the integration circle (cf.\ the discussion in 
ref.~\cite{pivsuppl}). One is not allowed to 
use integration contours that go close 
to the origin
because this region is completely nonperturbative and should be avoided:
perturbation theory cannot decipher the structure of amplitudes 
in this region pointwise, only contributions to
the integrals are perturbative and can be computed. This situation is
to some extend analogous to the situation with Coulombic poles
especially for not very heavy quarks. 
For perturbation theory applications any
type of infrared singularity should be 
avoided by moving the integration contour far from the origin
and keeping infrared nonperturbative region inside, 
thereby including also the contribution of this region into 
the integral.
The possibility to accurately apply perturbation theory for averaged 
quantities is a specific feature of two-point correlators with simple
analytic properties in the momentum variable. 
In the cases when observables are obtained by the averaging of 
more complicated Green's functions where the analytic 
structure is not transparent the effects of running are accounted for
by considering a model field theory with a one-loop gluon 
propagator reiterated in all orders of perturbation theory. 
To respect gauge invariance in QCD in such a model 
the technique of naive nonabelianization is
used \cite{be0}. Note that in pure gluodynamics which is a proper
theoretical model for studying glueballs this trick is not
straightforward. If analytic properties of the amplitude 
are unknown one has no clear way to avoid going through infrared 
singularities of the running coupling 
and one is trying to perform the integration across the infrared region
directly (as in applications of infrared renormalons~\cite{benren}). 
In this case an infrared structure of the running is important 
for the analysis, however, it is completely nonperturbative. 
Therefore, the obtained results depend on additional assumptions
about the infrared behavior of the running coupling.

As a last remark we give an expression for 
the resummed function $\Pi_G(q^2)$
in the second order of the $\beta$-function.
Taking the approximation for the $\beta$-function in the form
\begin{equation}
\label{betaF2}
\beta(a)\equiv \beta_2(a)=-a^2\left(\beta_0+\beta_1 a
\right)
\end{equation}
one finds the expression for the resummed function $\Pi_G(q^2)$:
\begin{equation}
\label{piRG2}
\Pi_G(Q^2)=\frac{\pi^2}{\beta_1}\ln\left(
\beta_0+\beta_1\frac{\alpha_G^{(2)}(Q^2)}{\pi}
\right)
+{\rm subtractions}
\end{equation}
where the function $\alpha_G^{(2)}(Q^2)$ is a solution of the 
renormalization group equation with the second
order $\beta$-function
\begin{equation}
\label{RG2}
Q^2\frac{d}{dQ^2}\pfrac{\alpha_G^{(2)}(Q^2)}{\pi}
=-\beta_0\pfrac{\alpha_G^{(2)}(Q^2)}{\pi}^2
-\beta_1\pfrac{\alpha_G^{(2)}(Q^2)}{\pi}^3\, . 
\end{equation}
The generalization of our analysis to this case is straightforward.

To conclude, it has been shown that for the observables related to 
two-point current correlators 
the summation of the effects of running 
can be done in perturbation theory.  
In more complicated cases without simple analytic structure
of the respective Green's functions
the interpretation of running in the infrared region is not unique
and is outside the scope of 
perturbation theory. The asymptotic structure of the 
perturbation theory series depends on 
the actual treatment of the observables 
(there is no true asymptotic structure unless explicit assumptions are
formulated). The series can be analytic at the origin for some
approximations as it is for the widely used approximation 
with resummation on the contour.
Possible power corrections stemming from such resummation 
are of rather computational origin and simply reflect a particular way of
approximating the relevant integrals; no general conclusions 
on the analytic structure of the exact theory can be drawn. 
Theoretically, there is no invariant meaning 
in splitting the results into nonperturbative 
infrared power corrections and perturbation theory
part (as opposite to OPE where power corrections 
are related to high-dimension operators and determined by 
the projections onto
other perturbation theory states than the vacuum). 
Phenomenologically,
the high-order perturbation theory terms (with high powers of inverse
logarithms) can numerically mimic the renormalon-type 
power corrections well. In this situation, 
the way to go beyond perturbation theory for improving the accuracy 
of theoretical formulae would be just a convention to use for the
observables an effective scheme where all perturbative
corrections are explicitly resummed into the redefinition of the coupling.

\subsection*{Acknowledgements}
The work is supported in part by the Russian Fund for Basic Research
under contracts 99-01-00091 and 01-02-16171. A.A.~Pivovarov is Alexander von
Humboldt fellow.

\end{document}